\documentclass[aps,pra,notitlepage,%
superscriptaddress,longbibliography,twocolumn]{revtex4-1}

\usepackage[colorlinks=true, allcolors=blue]{hyperref}
\usepackage{natbib}
\usepackage{graphicx}
\usepackage{amssymb}
\usepackage{amsmath}
\usepackage{ifthen}
\usepackage{braket}
\usepackage{xcolor}
\usepackage{bm}
\usepackage{bbm}
\usepackage{cancel}
\usepackage{comment}
\usepackage{siunitx}
\usepackage{float}
\usepackage{dcolumn}
\newcolumntype{d}[1]{D{.}{.}{#1}}
\newcolumntype{.}{D{x}{}{9}}
\newcolumntype{,}{D{x}{}{5}}
\newcolumntype{;}{D{x}{}{19}}
\usepackage{subfigure}
\usepackage{fancyvrb}
\usepackage{maybemath}
\usepackage{upgreek}

\newcommand{\dimmu}{\mbox{\sl\textmu}}

\newcommand{\MSbar}{\overline{\mbox{MS}}}

\definecolor{garrosgreen}{rgb}{0.1, 0.4, 0.1}
\definecolor{dartmouthgreen}{rgb}{0.05, 0.5, 0.06}
\definecolor{duelferred}{rgb}{0.7, 0.2, 0.1}
\definecolor{cambridgeblue}{rgb}{0.1, 0.3, 1.0}
\definecolor{oxfordblue}{rgb}{0.05, 0.2, 0.7}

\newcommand{\calH}{{\mathcal H}}
\newcommand{\calE}{{\mathcal E_n}}
\newcommand{\calL}{{\mathcal L}}
\newcommand{\calO}{{\mathcal O}}

\newcommand{\dd}{\mathrm{d}}
\newcommand{\ii}{\mathrm{i}}
\newcommand{\ee}{\mathrm{e}}

\newcommand{\vp}{\mathrm{vp}}

\usepackage{xcolor}
\usepackage{bbm}

\definecolor{light}{gray}{0.90}
\definecolor{darker}{gray}{0.50}
\definecolor{dark}{gray}{0.30}

\begin{document}

\title{Reexamination of vacuum-polarization corrections \\
to the self-energy in muonic bound systems}

\author{B. Ohayon}
\affiliation{The Helen Diller Quantum Center, Department of Physics,
Technion-Israel Institute of Technology, Haifa, 3200003, Israel}

\author{U. D. Jentschura}
\affiliation{Department of Physics and LAMOR,
Missouri University of Science and Technology,
Rolla, Missouri 65409, USA}

\begin{abstract}
In muonic bound systems, the dominant radiative correction is due to vacuum
polarization.  Yet, for the interpretation of  precision experiments,
self-energy  effects are also important.  In turn, additional
vacuum-polarization loops perturb the self-energy.  Here, we show that combined
self-energy vacuum-polarization effects can perturb the bound-state self-energy
at the percent level in one-muon bound systems with nuclear charges 
$Z = 1$---$6$.  We also update previous treatments of the corrections for muonic
hydrogen, muonic deuterium, and muonic helium bound systems.
\end{abstract}

\maketitle

\tableofcontents

%
%
\section{Introduction}

The one-loop electronic vacuum-polarization (eVP) correction to the Coulomb
potential constitutes the most numerically important radiative correction for
the energies of bound states in muonic bound systems.  Due to the smaller
effective Bohr radius as compared to electronic systems, the muon penetrates
the vacuum-polarized charge cloud around the nucleus  very effectively,
and the bound muon is exposed to a  mean electric field strength,
which, even for nuclear charge numbers $Z > 3$, surpasses not only the field
strength achievable in highly charged hydrogenlike ions with the heaviest
nuclei, but also, Schwinger's critical electric field strength (see Fig.~3 of
Ref.~\cite{Je2015muonic}).

Specifically, the leading energy shift mediated by eVP in muonic bound systems
is of the order of $\alpha (Z\alpha)^2 m_r$, where $m_r$ is the reduced mass of
the bound system, $Z$ is the nuclear charge, and $\alpha$ is the fine-structure
constant (we set $\hbar = c = \epsilon_0 = 1$).  Self-energy (SE) shifts, by
contrast, are parametrically suppressed in muonic bound
systems, and are of order $\alpha(Z\alpha)^4 m_r$,
enhanced by a logarithm of $Z\alpha$.
Except for the logarithmic enhancement,
this is the same magnitude as the correction due to vacuum
polarization with muon loops ($\mu$VP), as well as the first-order relativistic
correction to the eVP  (see Ref.~\cite{Pa1996muonic}).

\begin{figure}[t!]
\begin{center}
\begin{minipage}{0.99\linewidth}
\begin{center}
\includegraphics[width=0.91\linewidth]{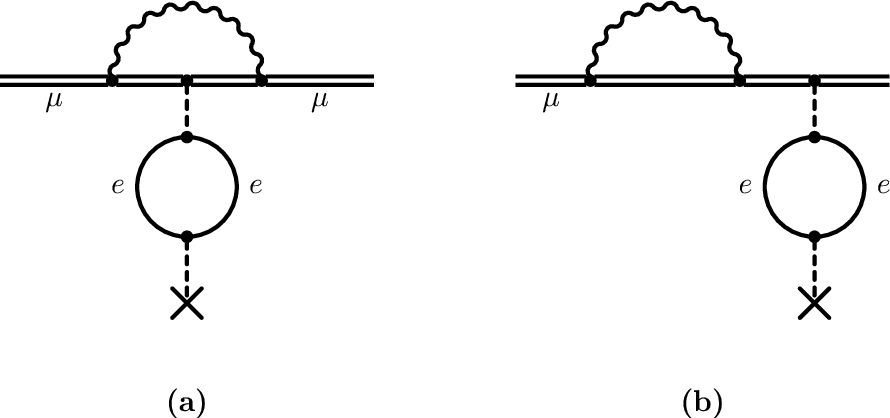}
\end{center}
\caption{\label{fig1} The Feynman diagrams 
for the vacuum-polarization correction
to the self-energy comprise a vacuum-polarization
insertion (a) into a Coulomb photon of the 
vertex diagram and the second-order effect (b).}
\end{minipage}
\end{center}
\end{figure}

\begin{figure}[t!]
\begin{center}
\begin{minipage}{0.99\linewidth}
\begin{center}
\includegraphics[width=0.91\linewidth]{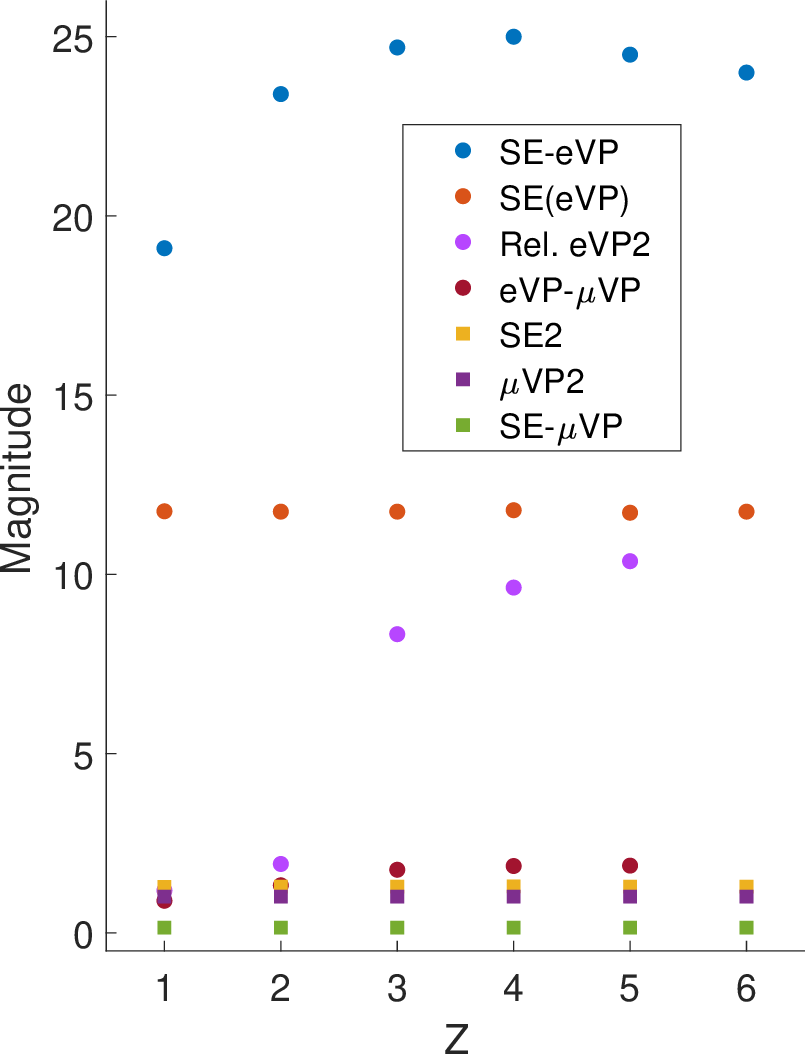}
\end{center}
\caption{\label{fig:cont} Magnitude of various contributions to the
$2P_{1/2}$--$2S$ Lamb shift in light muonic atoms, in units of
$(\frac{\alpha}{\pi})^2\frac{(Z\alpha)^4}{8}\frac{m_r^3}{m_\mu^2}$.
Contributions specific to muonic atoms are denoted with circles and include
the self-energy with electronic vacuum polarization (SE-eVP) calculated in this
work. The eVP insertion into the radiative photon
of the one-loop self energy is 
denoted as SE(eVP); it is taken from Eqs.~(65)---(68) from Ref.~\cite{PaEtAl2024}.
The relativistic contributions to two-loop eVP 
(Rel.~eVP2, without recoil corrections to that effect) 
are taken from Tables~II and~IV of
Ref.~\cite{KaIvKa2013}, and Tables~I---III of Ref.~\cite{KrMaMaSu2016}.
The one-loop eVP with one-loop $\mu$VP (eVP-$\mu$VP)
is taken from Table~VIII of Ref.~\cite{KaIvKa2013} and 
Tables~I-III of Ref.~\cite{KrMaMaSu2016}. 
From Table~3 in~\cite{YePaPa2019},
we can scale contributions originally obtained for electronic
systems, namely, the two-loop self-energy SE2, the two-loop
muonic vacuum polarization $\mu$VP2, and the combination of self-energy and
muonic vacuum polarization SE-$\mu$VP. These scaled contributions are 
denoted with squares.}
\end{minipage}
\end{center}
\end{figure}

In electronic systems, the next-to-leading correction to the 
SE is of order $\alpha(Z\alpha)^5 m_r$; it was calculated by
Baranger, Bethe, and Feynman~\cite{BaBeFe1953} (see also Chap.~15 of
Ref.~\cite{JeAd2022book}).  In muonic systems, by contrast, the
eVP-mediated binding correction to
the SE (SE-eVP), of which the corresponding
diagrams are shown in Fig.~\ref{fig1}, is logarithmically enhanced and of
order $\alpha^2 (Z\alpha)^4 \, \ln[(Z\alpha)^{-2}] m_r$; it is thus of
comparable magnitude to the effect of  order $\alpha (Z\alpha)^5 m_r$, if not
larger.  In light muonic atoms, this effect is the largest out
of all corrections which enter at $\alpha^2 (Z\alpha)^4 m_r$~\cite{KaIvKa2013},
as illustrated in Fig.~\ref{fig:cont}.

The SE-eVP effect in muonic atoms was first considered by Pachucki,
who calculated it in the leading logarithmic approximation [Eq.~(39)
in~\cite{Pa1996muonic}].  The numerical result for the $2P$--$2S$ Lamb shift in muonic
hydrogen was $-0.5~\mu$eV.  Eides {\em et al.}.~included it in their compilation,
with an estimated uncertainty of $20\%$~\cite{EiGrSh2001}. However, calculations
beyond the leading-log approximation have shown that there is a strong
cancelation between the leading logarithmic effect and the vacuum-polarization
correction  to the nonlogarithmic term~\cite{WuJe2011}, so that the
approximation only gives the correct order-of-magnitude.

Refined calculations of SE-eVP have been extended to muonic deuterium and
helium in Ref.~\cite{WuJe2011}. Together with many other calculated
corrections, they are currently used in extractions of nuclear charge radii
from the laser spectroscopy of the Lamb shift in muonic atoms (see
Ref.~\cite{PaEtAl2024} and references therein).  Notably, the size of the SE-eVP
correction in muonic hydrogen is of similar magnitude as the currently dominating uncertainty, which is that of the two-photon-exchange
correction~\cite{PaEtAl2024}.  Moreover, for elements with 
$Z>2$, a new generation of high-resolution x-ray spectroscopy 
experiments is underway~\cite{OhEtAl2024,UnEtAl2024}, 
which could be followed by laser-spectroscopic
studies~\cite{ScEtAl2018}. An interpretation of measured transitions in terms
of absolute nuclear charge radii necessitates the estimation of various QED
effects, including the SE-eVP, for which an accurate calculation is notably
missing in the literature.

In light of the above, in this work we validate and correct the current
results, increase their numerical precision, and extend them to heavier nuclei.
Natural units with $\hbar = c = \epsilon_0 = 1$ are used throughout.
In Sec.~\ref{sec2}, we consider a rederivation of the 
relevant expressions for the combined self-energy 
vacuum-polarization effect, based on a version of 
nonrelativistic quantum electrodynamics (NRQED) adapted 
to muonic bound systems.
An update on numerical results is presented in Sec.~\ref{sec3}.
Conclusions are reserved for Sec.~\ref{sec4}.

%
%
\section{Rederivation}
\label{sec2}

%
%
\subsection{NRQED Adapted to Muonic Systems}

We start from the Lagrangian of NRQED$_\mu$, 
which is a variant of NRQED (Chap.~17 of Ref.~\cite{JeAd2022book})
adapted to muonic bound systems~\cite{AdJe2024gauge},
\begin{align}
\label{NRQEDmu}
\mathcal{L} &=  \sum_i \psi_i^\dagger
\Bigl \{ \ii D_t + 
\frac{\vec D\, ^2}{2m_i} + \frac{\vec D\,^4}{8 m_i^3}
\cr
&+ c^{(i)}_F \frac{q_i}{2m_i} \vec \sigma_i \cdot \vec B 
\cr
&+ c^{(i)}_D \frac{q_i}{8 m_i^2}
\left ( \vec D \cdot \vec E - \vec E \cdot \vec D \right )
\cr
&+ c^{(i)}_S \frac{\ii q_i}{8 m_i^2} \vec \sigma_i \cdot
\left ( \vec D \times \vec E - \vec E \times \vec D \right ) 
+ \cdots \Bigr \} \psi_i 
\cr
&+ \bar \Psi_e \left ( \ii \, \gamma^\mu \, D_\mu - 
m_e \right ) \Psi_e 
\cr
&+ \mbox{four-fermion contact terms} \cr
&+ \mbox{photon terms} + \mathrm{counter terms} \cr
&+ \mbox{higher-order terms} \,.
\end{align}
The fields $\psi_i$ are explicit nonrelativistic
spinor fields, describing particles with charge $q_i$.
The fields $\vec E$ and $\vec B$ are the 
quantized electric and magnetic fields.
However, according to Eq.~(17.6) of 
Ref.~\cite{JeAd2022book}, one should note that $\vec E$ is given
by 
\begin{equation}
\vec E = -\vec \nabla A^0 - 
\frac{\partial \vec A}{\partial t} \,,
\end{equation}
where the scalar potential $A^0$ does not participate 
in the quantization in Coulomb gauge.
The covariant derivatives are defined via the relations
\begin{subequations}
\label{def_cov_derivative}
\begin{eqnarray}
D_t \, \psi_i &=& 
\left( \frac{\partial}{\partial t} + 
\ii \, q_i \, A^0 \right) \psi_i\,, \\
\vec D \, \psi_i &=& 
\left( \vec \nabla - \ii \, q_i \, \vec A \right) \, \psi_i \, .
\end{eqnarray}
\end{subequations}
In Eq.~\eqref{NRQEDmu}, $\Psi_e$ is the fully relativistic electron-positron
field operator. The necessity to 
include the electron-positron field in its full
relativistic form stems from the fact that 
the upper cutoff for the NRQED Lagrangian is 
at the order of the muon mass scale 
$m_\mu \sim 106 \, {\rm MeV}$. For ordinary 
QED, the upper cutoff is at the electron mass scale.

For NRQED$_\mu$, one needs to  take the electron-positron field into account
relativistically because the electron mass scale is far below the cutoff scale.
Therefore, in particular,  the integration of the virtual electron-positron
one-loop insertion into the photon propagator in NRQED$_\mu$ leads to the
Uehling potential in unexpanded form, as exemplified in the replacement
\begin{equation}
\label{repl}
V_C = e \, A^0 \to
V_C + V_\vp \,,
\end{equation}
where $V_C$ is the Coulomb potential.
In the following, we shall use the nuclear Coulomb potential
$V_C$ and the Uehling $V_\vp$ in the NRQED$_\mu$ Lagrangian within the 
external-field approximation.
The Uehling potential $V_\vp$ is recalled in Appendix~\ref{appa}, with
its definition given in Eq.~\eqref{Uvp}.
Within NRQED$_\mu$, the Uehling potential is perturbative in the
sense that the Coulomb potential is dominant,
and the Uehling correction is suppressed
by one power of the fine-structure constant.
So, one cannot absorb the electronic vacuum polarization
into a matching coefficient $c_{\rm VP}$. 
We remember that this matching coefficient
otherwise enters the NRQED Lagrangian 
via the replacement
$-(1/4) \, F_{\mu\nu} \, F^{\mu\nu} \to 
-(1/4) \, F_{\mu\nu} \, F^{\mu\nu} + 
c_{\rm VP} \, F_{\mu\nu} \, [ \partial^2 F^{\mu\nu} ]$,
where $c_{\rm VP} = \alpha/(15 \pi)$ 
according to Eq.~(17.29) of Ref.~\cite{JeAd2022book}.
For NRQED$_\mu$, the matching coefficient is appropriate for 
muonic vacuum polarization ($\mu$VP) but not eVP.

For our purposes, we can restrict the sum  over $i$ in Eq.~\eqref{NRQEDmu} to
the term $i = \mu$ for the (negatively charged) muon.  For the matching
coefficients, we obtain according to Eq.~(17.26) of Ref.~\cite{JeAd2022book},
to one-loop order,
\begin{subequations}
\label{matching_coeff}
\begin{align}
c^{(\mu)}_F =& \; 1 + \frac{\alpha}{2\pi} + \calO(\alpha^2) , \\
c^{(\mu)}_D =& \; 1 + \frac{4 \alpha}{3 \pi}
\left \{ -\frac{1}{\epsilon}
+ 2 \ln \left ( \frac{m_\mu}{\dimmu} \right ) \right \} + \calO(\alpha^2) , \\
c^{(\mu)}_S =& \; 1 + \frac{\alpha}{\pi} + \calO(\alpha^2).
\end{align}
\end{subequations}
Here, the number of dimensions is $d = 4 - 2 \epsilon$, while 
$\mbox{\dimmu}$ is the renormalization scale.

We consider the effects mediated by the Feynman diagrams in Fig.~\ref{fig1};
specifically, Fig.~\ref{fig1}(a) shows a vertex correction with an eVP
insertion in the Coulomb photon, while the diagram in Fig.~\ref{fig1}(b) is a
second-order effect involving both the bound-state self energy as well as the
vacuum-polarization.  The double line denotes the muon propagating in the
binding Coulomb field.  The cross denotes the interaction with the nucleus. For
low-energy virtual photons in the vacuum-polarization loop, diagram~(a)
represents the energy and Hamiltonian corrections to the Bethe logarithm. Both
diagrams were treated in Ref.~\cite{WuJe2011} and are revisited here.  The
SE-eVP correction can be split up into a high-energy and a low-energy part,
depending on the energy of the virtual photon.

%
%
\subsection{Low--Energy Part}
\label{LEP}

The inclusion of the fully relativistic electron-positron field operator in
Eq.~\eqref{NRQEDmu} implies that the photon propagator, in NRQED$_\mu$,
receives corrections due to eVP.  In Coulomb gauge, this implies that the
binding Coulomb potential $V_C$ receives a perturbative 
correction due to the Uehling potential $V_\vp$. 
For two-body bound systems, the
spirit of NRQED$_\mu$ implies that the Uehling potential needs to be included
in the potential energy $e \, A^0$ according to the replacement~\eqref{repl}.
We define the corrected Hamiltonian $\calH$, the corrected bound-state energy
$\calE$, the corrected bound-state wave function $\Psi_n$, and the correction
$\delta \phi_n$ of the bound-state wave function, as follows,
\begin{subequations}
\begin{align}
\calH =& \; H_S + V_\vp \,,
& \calE & = E_n + E_\vp \,,
\\[2ex]
\Psi_n =& \; \phi_n + \delta \phi_n \,,
& \delta \phi_n & = \frac{1}{(E_n - H_S)'} \, 
V_\vp \, | \phi_n \rangle \,.
\end{align}
\end{subequations}
Here, $[1/(E_n - H_S)']$ is the reduced Green function,
and $H_S$ denotes the unperturbed Schr\"{o}dinger Hamiltonian,
the Schr\"{o}dinger--Coulomb energy is $E_n$, and the unperturbed Schr\"{o}dinger--Pauli state is denoted as $\phi_n$, where $n$ is a multi-index summarizing the bound-state quantum numbers.

Under the addition of the Uehling  potential, the low-energy part of the self-energy, in dimensional regularization [see Eq.~(11.154) of Ref.~\cite{JeAd2022book}] generalizes to 
\begin{multline}
\label{ELEPstart}
\Delta E_{\rm LEP} =
e^2 \, \frac{D - 1}{D} \,
\frac{ \Omega_D }{ (2 \pi)^D } \, \frac12 \,
\int_\Lambda^\infty \dd k \, k^{D-2} 
\\[2ex]
\times
\left< \Psi_n \left| 
\frac{p^i}{m_\mu} \left( - \frac{1}{k} + \frac{\calH - \calE}{k^2} -
\frac{(\calH - \calE)^2}{k^3} \right) \frac{p^i}{m_\mu} 
\right| \Psi_n \right>
\\[2ex]
+ \; \frac{ 2 \alpha }{ 3 \pi } \,
\int_0^\Lambda \dd k \, k \,
\left< \Psi_n \left| 
\frac{p^i}{m_\mu} \frac{1}{\calE - k - \calH} \frac{p^i}{m_\mu} 
\right| \Psi_n \right> \,.
\end{multline}
Here, $D = 3 - 2\epsilon$ is the space dimension,
while $\Omega_D = 2 \pi^{D/2}/\Gamma(D/2)$
is the volume of the unit sphere embedded in $D$-dimensional space.
The idea is to introduce $\Lambda$ as a 
scale-separation parameter, which acts as an
infrared regulator for the first term in 
Eq.~\eqref{ELEPstart}, while it acts as 
an ultraviolet regulator for the second term 
in Eq.~\eqref{ELEPstart}. In view of the presence
of an infrared regulator in the first term 
of Eq.~\eqref{ELEPstart}, we can expand the 
integrand for large $\Lambda$. The regularization is done 
dimensionally, so we use the results
[see Eq.~(11.156) of Ref.~\cite{JeAd2022book}]
$\int_\Lambda^\infty {\dd}k \, k^{D-3} =
-\Lambda + \calO(\varepsilon)$,
and
$\int_\Lambda^\infty {\dd}k \, k^{D-4} =
\frac{1}{2 \varepsilon} - \ln(\Lambda) + \calO(\varepsilon)$.
Furthermore, one has the 
relation
$\int_\Lambda^\infty {\dd}k \, k^{D-5} =
\frac{1}{\Lambda} + {\cal O}(\varepsilon)$.
The last of these results implies that the term proportional
to $\frac{(\calH - \calE)^2}{k^3}$ in the 
integrand in Eq.~\eqref{ELEPstart} can be neglected in
the limit of large $\Lambda$. 
One can easily convince oneself that the dependence on 
$\Lambda$ cancels when both terms in Eq.~\eqref{ELEPstart} 
are added.

Using the $\MSbar$ conventions for the charge
(see Chap.~10 of Ref.~\cite{JeAd2022book}),
\begin{align}
\label{eDim}
e^2 &= (4 \pi)^{D/2-1/2} \, \alpha \,
\dimmu^{3 - D} \, \ee^{\gamma_E (3/2 - D/2) } \cr
&= (4 \pi)^{1 - \epsilon} \, \alpha \,
\dimmu^{2\epsilon} \ee^{\gamma_E \varepsilon} \,,
\end{align}
and the identity
\begin{equation}
e^2 \, \frac{D - 1}{D} \,
\frac{ \Omega_D }{ (2 \pi)^D } \frac12 =
\frac{2\alpha}{3\pi} + \frac{\alpha \epsilon}{\pi} \left[ \frac{10}{9} +
\frac43 \, \ln\left( \frac{ \dimmu }{2} \right)  \right] 
+ \calO(\epsilon^2)\,,
\end{equation}
one obtains, following 
Eq.~(11.157) of Ref.~\cite{JeAd2022book},
\begin{multline}
\label{CANCEL}
\Delta E_{\rm LEP} = \frac{\alpha}{\pi} \,
\left\{ \frac{2}{3} + \epsilon \,
\left[ \frac{10}{9} +
\frac43\, \ln\left(\frac{\dimmu}{2}\right) \right] \right\} 
\biggl[ \left( \frac{1}{2\epsilon} - \ln(\Lambda) \right) 
\\
\times \left< \Psi_n \left| \frac{p^i}{m_\mu} 
(\calH-\calE) \, \frac{p^i}{m_\mu} \right| \Psi_n \right> +
\Lambda \left< \Psi_n \left| \frac{\vec{p}^{\;2}}{m_\mu^2} 
\right| \Psi_n \right> \biggr]
\\[2ex]
+ \frac{4 \alpha}{3 \pi}\, \frac{(Z\alpha)^4}{n^3}
\frac{m_r^3}{m_\mu^2} \,
\ln\left(\frac{2 \Lambda}{(Z\alpha)^2 m_r}\right) 
\frac{\left< \Psi_n | \left[ p^i,
\left[  \calH - \calE, p^i \right] \right] | \Psi_n \right>}%
{4 (Z\alpha)^4 m_r^3} 
\\
- \frac{2 \alpha}{3 \pi}\,\Lambda \, 
\left< \Psi_n \left| \frac{\vec{p}^{\;2}}{m_\mu^2} \right| \Psi_n \right>
- \frac{4 \alpha}{3 \pi}\, \frac{(Z\alpha)^4}{n^3} 
\frac{m_r^3}{m_\mu^2} \,
\ln K_0(n, \ell) \,.
\end{multline}
Here, $\ln K_0(n, \ell)$ is the Bethe logarithm,
generalized to the potential $V_C + V_\vp$, which reads
\begin{multline}
\ln K_0(n \ell) = \frac{n^3}{2 (Z\alpha)^4}
\\
\times
\left< \Psi_n \left| \frac{p^i}{m_r} \,
\frac{\calH - \calE}{m_r}
\ln \left( \frac{2 \left| \calH - \calE \right|}{(Z\alpha)^2\,m_r} \right) \,
\frac{p^i}{m_r} \right| \Psi_n \right>\,.
\end{multline}
The dependence on $\Lambda$ in 
Eq.~\eqref{CANCEL} cancels, as expected,
and the low-energy part is obtained as follows,
\begin{multline}
\label{DeltaLEPcomplete}
\Delta E_{\rm LEP} = \frac{4 \alpha}{3 \pi} \,
\frac{(Z\alpha)^4}{n^3} \,
\frac{m_r^3}{m_\mu^2} \,
\biggl[ \left( \frac{1}{2 \epsilon} + \frac56 +
\ln\left( \frac{\mu}{ (Z\alpha)^2 m_r } \right) \right) \,
\\
\times 
\frac{\left< \Psi_n | \left[ p^i,
\left[  \calH - \calE, p^i \right] \right] | \Psi_n \right>}%
{4 (Z\alpha)^4 m_r^3} 
- \ln K_0(n, \ell) \biggr] \,.
\end{multline}
%
After a perturbative expansion in the 
vacuum-polarization potential, one obtains
\begin{multline}
\left< \Psi_n | \left[ p^i,
\left[  \calH - \calE, p^i \right] \right] | \Psi_n \right>
= \left< \phi_n \left| \vec\nabla^2( V_C ) 
\right| \phi_n \right> 
\\
+ \left< \phi_n  \left| \vec\nabla^2( V_\vp ) \right| \phi_n \right>
+ 2 \left< \phi_n \left|  \vec\nabla^2( V_C ) 
\right| \delta \phi_n \right> + \dots
\end{multline}
where the first term is the leading one,
pertinent to the Coulomb potential.
One verifies that, in leading order,
$\frac{\left< \Psi_n | \left[ p^i,
\left[  \calH - \calE, p^i \right] \right] | \Psi_n \right>}
{4 (Z\alpha)^4 m_r^3}
\approx
\frac{ \left< \phi_n  \left| \vec\nabla^2( V_C ) 
\right| \phi_n \right> }{4 (Z\alpha)^4 m_r^3}
= \delta_{\ell 0}/n^3$,
so that to the first perturbative order in $\alpha/\pi$, one has
\begin{equation}
\frac43 \, \frac{\left< \Psi_n | \left[ p^i,
\left[  \calH - \calE, p^i \right] \right] | \Psi_n \right>}%
{4 (Z\alpha)^4 m_r^3} =
\frac43 \, \frac{\delta_{\ell 0}}{n^3} +
\frac{\alpha}{\pi} \frac{V_{61}}{n^3}  \,, 
\end{equation}
where
\begin{equation}
V_{61} = \frac{4}{3}
\frac{\left< \phi_n \left| \vec\nabla^2( V_\vp ) \right| \phi_n \right>
+ 2 \left< \phi_n \left|  \vec\nabla^2( V_C ) 
\right| \delta \phi_n \right> }{ 4 (Z\alpha)^4 m_r^3/n^3 } \,.
\end{equation}
The Bethe logarithm can be expanded as follows,
\begin{equation}
\ln K_0(n \ell) = \ln k_0(n \ell) + 
\frac{\alpha}{\pi} \left( L^{(E)}_{60} + L^{(\psi)}_{60} + L^{(H)}_{60} \right) \,,
\end{equation}
where $\ln k_0(n, \ell)$ is the leading term,
obtained for the Coulomb potential.
We recall the relevant formula here in a
discrete-state representation,
\begin{multline}
\ln k_0(n \ell) = \frac{n^3}{2 (Z\alpha)^4} \sum_a
\left| \left< \phi_n \left| \frac{\vec p}{m_r} \right| \phi_a \right> \right|^2 \;
\frac{E_a - E_n}{m_r}
\\
\times \ln \left( \frac{2 \left| E_a - E_n \right|}{(Z\alpha)^2\,m_r} \right) \,,
\end{multline}
where the sum over $a$ contains the continuous spectrum.

\begin{widetext}
Specifically, the corrections are given by the following formulas.
For the energy-induced correction to the Bethe logarithm, one obtains
\begin{align}
\label{eq:LE}
\frac{\alpha}{\pi} L^{(E)}_{60} 
&= -\frac{n^3}{2 (Z\alpha)^4}
\left< \phi_n \left| \frac{p^i}{m_r} \, \left[ 1  +
\ln \left( \frac{2 \left| H_{\rm S} - E_n \right|}{(Z\alpha)^2\,m_r} \right)
\right]  \, \frac{p^i}{m_r} \right| \phi_n \right> 
\langle \phi_n | V_\vp | \phi_n \rangle 
\cr
&= -\frac{n^3}{2 (Z\alpha)^4} \sum_a
\left| \left< \phi_n \left| \frac{\vec p}{m_r} \right| \phi_a \right> \right|^2 \;
\left[ 1 + \ln \left( \frac{2 \left| E_a - E_n \right|}{(Z\alpha)^2\,m_r} \right) 
\right] \, \langle \phi_n | V_\vp | \phi_n \rangle \,.
\end{align}
The wave-function induced correction to the Bethe logarithm is given as follows:
\begin{align}
\label{eq:LP}
\frac{\alpha}{\pi} L^{(\psi)}_{60} =& \;
2 \times \frac{n^3}{2 (Z\alpha)^4}
\left< \phi_n \left| \frac{p^i}{m_r} \,
\frac{H_{\rm S} - E_n}{m_r}
\ln \left( \frac{2 \left| H_{\rm S} - E_n \right|}{(Z\alpha)^2\,m_r} \right) \,
\frac{p^i}{m_r} \left( \frac{1}{E_n - H_S} \right)' V_\vp 
\right| \phi_n \right>
\cr
=& \; \frac{n^3}{2 (Z\alpha)^4} \sum_{a} \sum_{b \neq n}
\left< \phi_n \left| \frac{\vec p}{m_r} \right| \phi_a \right> 
\frac{E_a - E_n}{m_r} \,
\ln \left( \frac{2 \left| E_a - E_n \right|}{(Z\alpha)^2\,m_r} \right) \,
\left< \phi_a \left| \frac{\vec p}{m_r} \right| \phi_b \right> \,
\frac{ \left< \phi_b \left| V_\vp \right| \phi_n \right> }{E_n - E_b} \,.
\end{align}
In these formulas, the vacuum-polarization induced correction to the wave
function $\delta \phi_n$ enters.
The Hamiltonian-induced correction to the Bethe logarithm is given as follows:
\begin{align}
\label{eq:LH}
\frac{\alpha}{\pi} L^{(H)}_{60} =& \;
\frac{n^3}{2 (Z\alpha)^4} \sum_{a} \sum_{b \neq a}
\frac{ \left< \phi_n \left| \frac{\vec p}{m_r} \right| \phi_a \right> 
\left< \phi_b \left| \frac{\vec p}{m_r} \right| \phi_n \right> }{E_a - E_b} \,
\left< \phi_a | V_\vp | \phi_b \right> \,
\biggl[ \frac{E_a - E_n}{m_r} \,
\ln \left( \frac{2 \left| E_a - E_n \right|}{(Z\alpha)^2\,m_r} \right) 
\nonumber\\
& \; - \frac{E_b - E_n}{m_r} \,
\ln \left( \frac{2 \left| E_b - E_n \right|}{(Z\alpha)^2\,m_r} \right) 
\biggr]
+ \frac{n^3}{2 (Z\alpha)^4} \sum_{a} 
\left| \left< \phi_n \left| \frac{\vec p}{m_r} \right| \phi_a \right> \right|^2 \,
\frac{\left< \phi_a | V_\vp | \phi_a \right>}{m_r} \,
\left[ 1 + 
\ln \left( \frac{2 \left| E_a - E_n \right|}{(Z\alpha)^2\,m_r} \right) 
\right] \,.
\end{align}
One might wonder about the 
contribution from $a=n$ to the second term
in the above expression, in view of the fact
that the expression
$\ln \left( \frac{2 \left| E_a - E_n \right|}{(Z\alpha)^2\,m_r} \right)$
diverges logarithmically when $E_a \to E_n$. 
However, when $| \phi_a \rangle = | \phi_n \rangle$,
the dipole transition matrix element
$\left< \phi_n \left| \frac{\vec p}{m_r} \right| \phi_a \right>$ vanishes
because of dipole selection rules.
Conversely, when $| \phi_a \rangle$ is energetically 
degenerate with $| \phi_n \rangle$ and 
a dipole transition is allowed, one can still show that 
the contribution to the second term in Eq.~\eqref{eq:LH} 
vanishes, in view of the fact the commutator relation
$\vec p/m_r = \ii [H_s, \vec r]$,
where $\vec r$ is the position operator for the 
muon-nucleus distance.
Summing Eqs.~\eqref{eq:LE}---\eqref{eq:LH}, the 
entire low-energy part is finally found as follows,
\begin{multline}
\label{ELEPres}
E_{\rm LEP} = \frac{4 \alpha}{3 \pi} \,
\frac{(Z\alpha)^4}{n^3} \,
\frac{m_r^3}{m_\mu^2} \,
\biggl[ \left( \frac{1}{2 \epsilon} + \frac56 +
\ln\left( \frac{\dimmu}{ (Z\alpha)^2 m_r } \right) \right) \,
\delta_{\ell 0} - \ln k_0(n, \ell) \biggr] 
\\
+ \left( \frac{\alpha}{\pi} \right)^2 \,
\frac{(Z\alpha)^4}{n^3} \,
\frac{m_r^3}{m_\mu^2} \,
\biggl[ \left( \frac{1}{2 \epsilon} + \frac56 +
\ln\left( \frac{\dimmu}{ (Z\alpha)^2 m_r } \right) \right) \,
V_{61} - \frac43 \, L_{60} \biggr] 
\,,
\qquad
L_{60} = L^{(E)}_{60} + L^{(\psi)}_{60} + L^{(H)}_{60} \,.
\end{multline}
\end{widetext}

An example is instructive.
For the $2S$ state of $\mu{}^{12}{\rm C}$, one obtains
$L_{60}^{(E)} = 0.042872$,
$L_{60}^{(\psi)} =  25.176577$,
$L_{60}^{(H)} =  -1.444258$,
so that the total is $L_{60} = 23.775190$.
The contribution of the wave function correction
is by far the numerically dominant term.
For the $2P$ state of $\mu{}^{12}{\rm C}$, one obtains
$L_{60}^{(E)} = -0.004020$,
$L_{60}^{(\psi)} = 0.133101$,
$L_{60}^{(H)} = -0.213113$,
and $L_{60} = -0.084033$.
For the $2P$ state, the contribution of the wave function correction
is not numerically dominant.

\begingroup
\squeezetable
\def\arraystretch{1.1}
\begin{table*}
\caption{\label{tableV61} $V_{61}$ are given for muonic bound systems
with nuclear charges $1 \leq Z \leq 6$. In the table,
$\mu$ denotes the bound muon in the one-muon ion (e.g., 
$\mu$H stands for muonic hydrogen).}
\begin{tabular}{c@{\hspace{5ex}}c@{\hspace{5ex}}l@{\hspace{5ex}}%
l@{\hspace{5ex}}l@{\hspace{5ex}}l}
\hline
\hline
 \multicolumn{1}{c}{Bound System} & $n$
 & \multicolumn{1}{c}{$nS$} & \multicolumn{1}{c}{$nP$}
 & \multicolumn{1}{c}{$nD$} & \multicolumn{1}{c}{$nF$} \\
\hline
\hline
\multicolumn{1}{l}{$\mu$H} &
  $n = 1$ & $  3.48559 \times 10^{0} $ & \multicolumn{1}{c}{---} & \multicolumn{1}{c}{---} & \multicolumn{1}{c}{---}  \\
& $n = 2$ & $  3.08601 \times 10^{0} $ & $ -2.32702 \times 10^{-2} $ & \multicolumn{1}{c}{---} & \multicolumn{1}{c}{---} \\
& $n = 3$ & $  2.97846 \times 10^{0} $ & $ -2.63586 \times 10^{-2} $ & $ -3.95816 \times 10^{-4} $ & \multicolumn{1}{c}{---} \\
& $n = 4$ & $  2.92701 \times 10^{0} $ & $ -2.73887 \times 10^{-2} $ & $ -5.39105 \times 10^{-4} $ & $ -5.10943 \times 10^{-6} $ \\
\hline
\multicolumn{1}{l}{$\mu$D} &
  $n = 1$ & $  3.61551 \times 10^{0} $ & \multicolumn{1}{c}{---} & \multicolumn{1}{c}{---} & \multicolumn{1}{c}{---} \\
& $n = 2$ & $  3.18785 \times 10^{0} $ & $ -2.45317 \times 10^{-2} $ & \multicolumn{1}{c}{---} & \multicolumn{1}{c}{---} \\
& $n = 3$ & $  3.07361 \times 10^{0} $ & $ -2.77042 \times 10^{-2} $ & $ -4.52469 \times 10^{-4} $ & \multicolumn{1}{c}{---} \\
& $n = 4$ & $  3.01904 \times 10^{0} $ & $ -2.87606 \times 10^{-2} $ & $ -6.14095 \times 10^{-4} $ & $ -6.39300 \times 10^{-6} $ \\
\hline
\multicolumn{1}{l}{${}^3$He} &
$n = 1$ & $  5.63331 \times 10^{0} $ & \multicolumn{1}{c}{---} & \multicolumn{1}{c}{---} & \multicolumn{1}{c}{---} \\
& $n = 2$ & $  4.71873 \times 10^{0} $ & $ -4.34521 \times 10^{-2} $ & \multicolumn{1}{c}{---} & \multicolumn{1}{c}{---} \\
& $n = 3$ & $  4.47997 \times 10^{0} $ & $ -4.66434 \times 10^{-2} $ & $ -2.13808 \times 10^{-3} $ & \multicolumn{1}{c}{---} \\
& $n = 4$ & $  4.36864 \times 10^{0} $ & $ -4.77518 \times 10^{-2} $ & $ -2.69903 \times 10^{-3} $ & $ -9.47682 \times 10^{-5} $ \\
\hline
\multicolumn{1}{l}{${}^4$He} &
$n = 1$ & $  5.66116 \times 10^{0} $ & \multicolumn{1}{c}{---} & \multicolumn{1}{c}{---} & \multicolumn{1}{c}{---} \\
& $n = 2$ & $  4.73968 \times 10^{0} $ & $ -4.36902 \times 10^{-2} $ & \multicolumn{1}{c}{---} & \multicolumn{1}{c}{---} \\
& $n = 3$ & $  4.49884 \times 10^{0} $ & $ -4.68662 \times 10^{-2} $ & $ -2.17286 \times 10^{-3} $ & \multicolumn{1}{c}{---} \\
& $n = 4$ & $  4.38660 \times 10^{0} $ & $ -4.79716 \times 10^{-2} $ & $ -2.73963 \times 10^{-3} $ & $ -9.75778 \times 10^{-5} $ \\
\hline
\multicolumn{1}{l}{${}^6$Li} &
$n = 1$ & $  6.99308 \times 10^{0} $ & \multicolumn{1}{c}{---} & \multicolumn{1}{c}{---} & \multicolumn{1}{c}{---}  \\
& $n = 2$ & $  5.75798 \times 10^{0} $ & $ -5.38775 \times 10^{-2} $ & \multicolumn{1}{c}{---} & \multicolumn{1}{c}{---} \\
& $n = 3$ & $  5.39966 \times 10^{0} $ & $ -5.60455 \times 10^{-2} $ & $ -4.16289 \times 10^{-3} $ & \multicolumn{1}{c}{---} \\
& $n = 4$ & $  5.23866 \times 10^{0} $ & $ -5.69489 \times 10^{-2} $ & $ -4.93839 \times 10^{-3} $ & $ -3.26874 \times 10^{-4}$ \\
\hline
\multicolumn{1}{l}{${}^{7}$Li} &
$n = 1$ & $  7.00182 \times 10^{0} $ & \multicolumn{1}{c}{---} & \multicolumn{1}{c}{---} & \multicolumn{1}{c}{---} \\
& $n = 2$ & $  5.76482 \times 10^{0} $ & $ -5.39360 \times 10^{-2} $ & \multicolumn{1}{c}{---} & \multicolumn{1}{c}{---} \\
& $n = 3$ & $  5.40560 \times 10^{0} $ & $ -5.60964 \times 10^{-2} $ & $ -4.17775 \times 10^{-3} $ & \multicolumn{1}{c}{---} \\
& $n = 4$ & $  5.24424 \times 10^{0} $ & $ -5.69982 \times 10^{-2} $ & $ -4.95397 \times 10^{-3} $ & $ -3.29113 \times 10^{-4} $ \\
\hline
 \multicolumn{1}{l}{${}^{9}$Be} &
$n = 1$ & $  7.97809 \times 10^{0} $ & \multicolumn{1}{c}{---} & \multicolumn{1}{c}{---} & \multicolumn{1}{c}{---} \\
& $n = 2$ & $  6.54468 \times 10^{0} $ & $ -5.97604 \times 10^{-2} $ & \multicolumn{1}{c}{---} & \multicolumn{1}{c}{---} \\
& $n = 3$ & $  6.07555 \times 10^{0} $ & $ -6.10772 \times 10^{-2} $ & $ -5.92496 \times 10^{-3} $ & \multicolumn{1}{c}{---} \\
& $n = 4$ & $  5.86969 \times 10^{0} $ & $ -6.17936 \times 10^{-2} $ & $ -6.71458 \times 10^{-3} $ & $ -6.51908 \times 10^{-4} $ \\
\hline
\multicolumn{1}{l}{${}^{10}$Be} &
$n = 1$ & $  7.98232 \times 10^{0} $ & \multicolumn{1}{c}{---} & \multicolumn{1}{c}{---} & \multicolumn{1}{c}{---} \\
& $n = 2$ & $  6.54813 \times 10^{0} $ & $ -5.97826 \times 10^{-2} $ & \multicolumn{1}{c}{---} & \multicolumn{1}{c}{---} \\
& $n = 3$ & $  6.07849 \times 10^{0} $ & $ -6.10959 \times 10^{-2} $ & $ -5.93277 \times 10^{-3} $ & \multicolumn{1}{c}{---} \\
& $n = 4$ & $  5.87242 \times 10^{0} $ & $ -6.18115 \times 10^{-2} $ & $ -6.72215 \times 10^{-3} $ & $ -6.53629 \times 10^{-4} $ \\
\hline
\multicolumn{1}{l}{${}^{10}$B} &
$n = 1$ & $  8.74561 \times 10^{0} $ & \multicolumn{1}{c}{---} & \multicolumn{1}{c}{---} & \multicolumn{1}{c}{---} \\
& $n = 2$ & $  7.18337 \times 10^{0} $ & $ -6.33768 \times 10^{-2} $ & \multicolumn{1}{c}{---} & \multicolumn{1}{c}{---} \\
& $n = 3$ & $  6.61634 \times 10^{0} $ & $ -6.41199 \times 10^{-2} $ & $ -7.34631 \times 10^{-3} $ & \multicolumn{1}{c}{---} \\
& $n = 4$ & $  6.36808 \times 10^{0} $ & $ -6.46835 \times 10^{-2} $ & $ -8.05589 \times 10^{-3} $ & $ -1.01009 \times 10^{-3} $ \\
\hline
\multicolumn{1}{l}{${}^{11}$B} &
$n = 1$ & $  8.74911 \times 10^{0} $ & \multicolumn{1}{c}{---} & \multicolumn{1}{c}{---} & \multicolumn{1}{c}{---} \\
& $n = 2$ & $  7.18633 \times 10^{0} $ & $ -6.33915 \times 10^{-2} $ & \multicolumn{1}{c}{---} & \multicolumn{1}{c}{---} \\
& $n = 3$ & $  6.61884 \times 10^{0} $ & $ -6.41323 \times 10^{-2} $ & $ -7.35273 \times 10^{-3} $ & \multicolumn{1}{c}{---} \\
& $n = 4$ & $  6.37036 \times 10^{0} $ & $ -6.46951 \times 10^{-2} $ & $ -8.06179 \times 10^{-3} $ & $ -1.01192 \times 10^{-3} $ \\
\hline
\multicolumn{1}{l}{${}^{12}$C} &
$n = 1$ & $  9.38273 \times 10^{0} $ & \multicolumn{1}{c}{---} & \multicolumn{1}{c}{---} & \multicolumn{1}{c}{---} \\
& $n = 2$ & $  7.73120 \times 10^{0} $ & $ -6.57915 \times 10^{-2} $ & \multicolumn{1}{c}{---} & \multicolumn{1}{c}{---} \\
& $n = 3$ & $  7.07880 \times 10^{0} $ & $ -6.61595 \times 10^{-2} $ & $ -8.49235 \times 10^{-3} $ & \multicolumn{1}{c}{---} \\
& $n = 4$ & $  6.78910 \times 10^{0} $ & $ -6.65964 \times 10^{-2} $ & $ -9.09077 \times 10^{-3} $ & $ -1.37077 \times 10^{-3} $ \\
\hline
\multicolumn{1}{l}{${}^{13}$C} &
$n = 1$ & $  9.38524 \times 10^{0} $ & \multicolumn{1}{c}{---} & \multicolumn{1}{c}{---} & \multicolumn{1}{c}{---} \\
& $n = 2$ & $  7.73339 \times 10^{0} $ & $ -6.58001 \times 10^{-2} $ & \multicolumn{1}{c}{---} & \multicolumn{1}{c}{---} \\
& $n = 3$ & $  7.08065 \times 10^{0} $ & $ -6.61667 \times 10^{-2} $ & $ -8.49675 \times 10^{-3} $ & \multicolumn{1}{c}{---} \\
& $n = 4$ & $  6.79078 \times 10^{0} $ & $ -6.66032 \times 10^{-2} $ & $ -9.09467 \times 10^{-3} $ & $ -1.37229 \times 10^{-3} $ \\
\hline
\hline
\end{tabular}
\end{table*}
\endgroup

\begingroup
\begin{table*}
\caption{\label{tableN60} Same as Table~\ref{tableV61}, but for $N_{60}$ Coefficients.}
\begin{tabular}{c@{\hspace{5ex}}c@{\hspace{5ex}}l@{\hspace{5ex}}%
l@{\hspace{5ex}}l}
\hline
\hline
 \multicolumn{1}{c}{Bound System} & $n$
 & \multicolumn{1}{c}{$nP$} & \multicolumn{1}{c}{$nD$} & \multicolumn{1}{c}{$nF$} \\
\hline
\multicolumn{1}{l}{$\mu$H} &
$n = 2$ & $  2.13804 \times 10^{-2} $ & \multicolumn{1}{c}{---} & \multicolumn{1}{c}{---} \\
& $n = 3$ & $  2.37590 \times 10^{-2} $ & $  1.54906 \times 10^{-4} $ & \multicolumn{1}{c}{---} \\
& $n = 4$ & $  2.44563 \times 10^{-2} $ & $  2.10042 \times 10^{-4} $ & $  1.16747 \times 10^{-6} $ \\
\hline
\multicolumn{1}{l}{$\mu$D} &
$n = 2$ & $  2.32830 \times 10^{-2} $ & \multicolumn{1}{c}{---} & \multicolumn{1}{c}{---} \\
& $n = 3$ & $  2.57550 \times 10^{-2} $ & $  1.82541 \times 10^{-4} $ & \multicolumn{1}{c}{---} \\
& $n = 4$ & $  2.64692 \times 10^{-2} $ & $  2.46521 \times 10^{-4} $ & $  1.50051 \times 10^{-6} $ \\
\hline
\multicolumn{1}{l}{$\mu^3$He} &
$n = 2$ & $  6.69813 \times 10^{-2} $ & \multicolumn{1}{c}{---} & \multicolumn{1}{c}{---} \\
& $n = 3$ & $  6.81191 \times 10^{-2} $ & $  1.40225 \times 10^{-3} $ & \multicolumn{1}{c}{---} \\
& $n = 4$ & $  6.82046 \times 10^{-2} $ & $  1.73793 \times 10^{-3} $ & $  3.48674 \times 10^{-5} $ \\
\hline
\multicolumn{1}{l}{$\mu^4$He} &
$n = 2$ & $  6.77811 \times 10^{-2} $ & \multicolumn{1}{c}{---} & \multicolumn{1}{c}{---} \\
& $n = 3$ & $  6.88460 \times 10^{-2} $ & $  1.43480 \times 10^{-3} $ & \multicolumn{1}{c}{---} \\
& $n = 4$ & $  6.89077 \times 10^{-2} $ & $  1.77565 \times 10^{-3} $ & $  3.61374 \times 10^{-5} $ \\
\hline
\multicolumn{1}{l}{$\mu^6$Li} &
$n = 2$ & $  1.12317 \times 10^{-1} $ & \multicolumn{1}{c}{---} & \multicolumn{1}{c}{---} \\
& $n = 3$ & $  1.07604 \times 10^{-1} $ & $  3.80629 \times 10^{-3} $ & \multicolumn{1}{c}{---} \\
& $n = 4$ & $  1.05919 \times 10^{-1} $ & $  4.36073 \times 10^{-3} $ & $  1.66880 \times 10^{-4} $ \\
\hline
\multicolumn{1}{l}{$\mu^7$Li} &
$n = 2$ & $  1.12649 \times 10^{-1} $ & \multicolumn{1}{c}{---} & \multicolumn{1}{c}{---} \\
& $n = 3$ & $  1.07883 \times 10^{-1} $ & $  3.82800 \times 10^{-3} $ & \multicolumn{1}{c}{---} \\
& $n = 4$ & $  1.06182 \times 10^{-1} $ & $  4.38324 \times 10^{-3} $ & $  1.68383 \times 10^{-4} $ \\
\hline
\multicolumn{1}{l}{$\mu^{9}$Be} &
$n = 2$ & $  1.52741 \times 10^{-1} $ & \multicolumn{1}{c}{---} & \multicolumn{1}{c}{---} \\
& $n = 3$ & $  1.40987 \times 10^{-1} $ & $  6.86843 \times 10^{-3} $ & \multicolumn{1}{c}{---} \\
& $n = 4$ & $  1.37108 \times 10^{-1} $ & $  7.40339 \times 10^{-3} $ & $  4.24422 \times 10^{-4} $ \\
\hline
\multicolumn{1}{l}{$\mu^{10}$Be} &
$n = 2$ & $  1.52927 \times 10^{-1} $ & \multicolumn{1}{c}{---} & \multicolumn{1}{c}{---} \\
& $n = 3$ & $  1.41138 \times 10^{-1} $ & $  6.88440 \times 10^{-3} $ & \multicolumn{1}{c}{---} \\
& $n = 4$ & $  1.37248 \times 10^{-1} $ & $  7.41868 \times 10^{-3} $ & $  4.25988 \times 10^{-4} $ \\
\hline
\multicolumn{1}{l}{$\mu^{10}$B} &
$n = 2$ & $  1.88102 \times 10^{-1} $ & \multicolumn{1}{c}{---} & \multicolumn{1}{c}{---} \\
& $n = 3$ & $  1.69680 \times 10^{-1} $ & $  1.02037 \times 10^{-2} $ & \multicolumn{1}{c}{---} \\
& $n = 4$ & $  1.63409 \times 10^{-1} $ & $  1.05066 \times 10^{-2} $ & $  7.95348 \times 10^{-4} $ \\
\hline
\multicolumn{1}{l}{$\mu^{11}$B} &
$n = 2$ & $  1.88270 \times 10^{-1} $ & \multicolumn{1}{c}{---} & \multicolumn{1}{c}{---} \\
& $n = 3$ & $  1.69817 \times 10^{-1} $ & $  1.02210 \times 10^{-2} $ & \multicolumn{1}{c}{---} \\
& $n = 4$ & $  1.63533 \times 10^{-1} $ & $  1.05223 \times 10^{-2} $ & $  7.97480 \times 10^{-4} $ \\
\hline
\multicolumn{1}{l}{$\mu^{12}$C} &
$n = 2$ & $  2.19659 \times 10^{-1} $ & \multicolumn{1}{c}{---} & \multicolumn{1}{c}{---} \\
& $n = 3$ & $  1.95305 \times 10^{-1} $ & $  1.36546 \times 10^{-2} $ & \multicolumn{1}{c}{---} \\
& $n = 4$ & $  1.86517 \times 10^{-1} $ & $  1.35812 \times 10^{-2} $ & $  1.26289 \times 10^{-3} $ \\
\hline
\multicolumn{1}{l}{$\mu^{13}$C} &
$n = 2$ & $  2.19786 \times 10^{-1} $ & \multicolumn{1}{c}{---} & \multicolumn{1}{c}{---} \\
& $n = 3$ & $  1.95410 \times 10^{-1} $ & $  1.36694 \times 10^{-2} $ & \multicolumn{1}{c}{---} \\
& $n = 4$ & $  1.86610 \times 10^{-1} $ & $  1.35942 \times 10^{-2} $ & $  1.26507 \times 10^{-3} $ \\
\hline
\hline
\end{tabular}
\end{table*}
\endgroup

\begingroup
\squeezetable
\def\arraystretch{1.1}
\begin{table*}
\caption{\label{tableL60} Same as Table~\ref{tableV61}, but for $L_{60}$ Coefficients.}
\begin{tabular}{c@{\hspace{5ex}}c@{\hspace{5ex}}l@{\hspace{5ex}}%
l@{\hspace{5ex}}l@{\hspace{5ex}}l}
\hline
\hline
 \multicolumn{1}{c}{Bound System} & $n$
 & \multicolumn{1}{c}{$nS$} & \multicolumn{1}{c}{$nP$}
 & \multicolumn{1}{c}{$nD$} & \multicolumn{1}{c}{$nF$} \\
\hline
\multicolumn{1}{l}{$\mu$H} &
$n = 1$ & $ 1.23958 \times 10^{1} $ & \multicolumn{1}{c}{---} & \multicolumn{1}{c}{---} & \multicolumn{1}{c}{---} \\
& $n = 2$ & $  1.12836 \times 10^{1} $ & $ -1.45598 \times 10^{-2} $ & \multicolumn{1}{c}{---} & \multicolumn{1}{c}{---} \\
& $n = 3$ & $  1.10260 \times 10^{1} $ & $ -1.75766 \times 10^{-2} $ & $  3.23884 \times 10^{-4} $ & \multicolumn{1}{c}{---} \\
& $n = 4$ & $  1.09110 \times 10^{1} $ & $ -1.89727 \times 10^{-2} $ & $  4.46707 \times 10^{-4} $ & $  8.50224 \times 10^{-6} $ \\
\hline
\multicolumn{1}{l}{$\mu$D} &
$n = 1$ & $  1.27625 \times 10^{1} $ & \multicolumn{1}{c}{---} & \multicolumn{1}{c}{---} & \multicolumn{1}{c}{---} \\
& $n = 2$ & $  1.15765 \times 10^{1} $ & $ -1.54461 \times 10^{-2} $ & \multicolumn{1}{c}{---} & \multicolumn{1}{c}{---} \\
& $n = 3$ & $  1.13033 \times 10^{1} $ & $ -1.86690 \times 10^{-2} $ & $  3.70482 \times 10^{-4} $ & \multicolumn{1}{c}{---} \\
& $n = 4$ & $  1.11815 \times 10^{1} $ & $ -2.01472 \times 10^{-2} $ & $  5.08205 \times 10^{-4} $ & $  1.06413 \times 10^{-5} $ \\
\hline
\multicolumn{1}{l}{$\mu^3$He} &
$n = 1$ & $  1.82976 \times 10^{1} $ & \multicolumn{1}{c}{---} & \multicolumn{1}{c}{---} & \multicolumn{1}{c}{---} \\
& $n = 2$ & $  1.58488 \times 10^{1} $ & $ -3.22934 \times 10^{-2} $ & \multicolumn{1}{c}{---} & \multicolumn{1}{c}{---} \\
& $n = 3$ & $  1.52902 \times 10^{1} $ & $ -3.93437 \times 10^{-2} $ & $  1.75344 \times 10^{-3} $ & \multicolumn{1}{c}{---} \\
& $n = 4$ & $  1.50450 \times 10^{1} $ & $ -4.20972 \times 10^{-2} $ & $  2.13524 \times 10^{-3} $ & $  1.59537 \times 10^{-4} $ \\
\hline
\multicolumn{1}{l}{$\mu^4$He} &
$n = 1$ & $  1.83727 \times 10^{1} $ & \multicolumn{1}{c}{---} & \multicolumn{1}{c}{---} & \multicolumn{1}{c}{---} \\
& $n = 2$ & $  1.59060 \times 10^{1} $ & $ -3.25737 \times 10^{-2} $ & \multicolumn{1}{c}{---} & \multicolumn{1}{c}{---} \\
& $n = 3$ & $  1.53426 \times 10^{1} $ & $ -3.96804 \times 10^{-2} $ & $  1.78155 \times 10^{-3} $ & \multicolumn{1}{c}{---} \\
& $n = 4$ & $  1.50954 \times 10^{1} $ & $ -4.24497 \times 10^{-2} $ & $  2.16492 \times 10^{-3} $ & $  1.64306 \times 10^{-4} $ \\
\hline
\multicolumn{1}{l}{$\mu^{6}$Li} &
$n = 1$ & $  2.19297 \times 10^{1} $ & \multicolumn{1}{c}{---} & \multicolumn{1}{c}{---} & \multicolumn{1}{c}{---} \\
& $n = 2$ & $  1.86418 \times 10^{1} $ & $ -4.77365 \times 10^{-2} $ & \multicolumn{1}{c}{---} & \multicolumn{1}{c}{---} \\
& $n = 3$ & $  1.78130 \times 10^{1} $ & $ -5.73498 \times 10^{-2} $ & $  3.31549 \times 10^{-3} $ & \multicolumn{1}{c}{---} \\
& $n = 4$ & $  1.74602 \times 10^{1} $ & $ -6.07326 \times 10^{-2} $ & $  3.59651 \times 10^{-3} $ & $  5.53860 \times 10^{-4} $ \\
\hline
\multicolumn{1}{l}{$\mu^{7}$Li} &
$n = 1$ & $  2.19528 \times 10^{1} $ & \multicolumn{1}{c}{---} & \multicolumn{1}{c}{---} & \multicolumn{1}{c}{---} \\
& $n = 2$ & $  1.86599 \times 10^{1} $ & $ -4.78479 \times 10^{-2} $ & \multicolumn{1}{c}{---} & \multicolumn{1}{c}{---} \\
& $n = 3$ & $  1.78291 \times 10^{1} $ & $ -5.74758 \times 10^{-2} $ & $  3.32625 \times 10^{-3} $ & \multicolumn{1}{c}{---} \\
& $n = 4$ & $  1.74755 \times 10^{1} $ & $ -6.08615 \times 10^{-2} $ & $  3.60521 \times 10^{-3} $ & $  5.57659 \times 10^{-4} $ \\
\hline
\multicolumn{1}{l}{$\mu^{9}$Be} &
$n = 1$ & $  2.45362 \times 10^{1} $ & \multicolumn{1}{c}{---} & \multicolumn{1}{c}{---} & \multicolumn{1}{c}{---} \\
& $n = 2$ & $  2.07107 \times 10^{1} $ & $ -6.13071 \times 10^{-2} $ & \multicolumn{1}{c}{---} & \multicolumn{1}{c}{---} \\
& $n = 3$ & $  1.96329 \times 10^{1} $ & $ -7.23762 \times 10^{-2} $ & $  4.49031 \times 10^{-3} $ & \multicolumn{1}{c}{---} \\
& $n = 4$ & $  1.91833 \times 10^{1} $ & $ -7.59775 \times 10^{-2} $ & $  4.41305 \times 10^{-3} $ & $  1.10344 \times 10^{-3} $ \\
\hline
\multicolumn{1}{l}{$\mu^{10}$Be} &
$n = 1$ & $  2.45473 \times 10^{1} $ & \multicolumn{1}{c}{---} & \multicolumn{1}{c}{---} & \multicolumn{1}{c}{---} \\
& $n = 2$ & $  2.07197 \times 10^{1} $ & $ -6.13697 \times 10^{-2} $ & \multicolumn{1}{c}{---} & \multicolumn{1}{c}{---} \\
& $n = 3$ & $  1.96408 \times 10^{1} $ & $ -7.24443 \times 10^{-2} $ & $  4.49498 \times 10^{-3} $ & \multicolumn{1}{c}{---} \\
& $n = 4$ & $  1.91907 \times 10^{1} $ & $ -7.60460 \times 10^{-2} $ & $  4.41566 \times 10^{-3} $ & $  1.10634 \times 10^{-3} $ \\
\hline
\multicolumn{1}{l}{$\mu^{10}$B} &
$n = 1$ & $  2.65584 \times 10^{1} $ & \multicolumn{1}{c}{---} & \multicolumn{1}{c}{---} & \multicolumn{1}{c}{---} \\
& $n = 2$ & $  2.23674 \times 10^{1} $ & $ -7.32449 \times 10^{-2} $ & \multicolumn{1}{c}{---} & \multicolumn{1}{c}{---} \\
& $n = 3$ & $  2.10699 \times 10^{1} $ & $ -8.52266 \times 10^{-2} $ & $  5.23394 \times 10^{-3} $ & \multicolumn{1}{c}{---} \\
& $n = 4$ & $  2.05288 \times 10^{1} $ & $ -8.88038 \times 10^{-2} $ & $  4.72294 \times 10^{-3} $ & $  1.69929 \times 10^{-3} $ \\
\hline
\multicolumn{1}{l}{$\mu^{11}$B} &
$n = 1$ & $  2.65676 \times 10^{1} $ & \multicolumn{1}{c}{---} & \multicolumn{1}{c}{---} & \multicolumn{1}{c}{---} \\
& $n = 2$ & $  2.23751 \times 10^{1} $ & $ -7.33019 \times 10^{-2} $ & \multicolumn{1}{c}{---} & \multicolumn{1}{c}{---} \\
& $n = 3$ & $  2.10765 \times 10^{1} $ & $ -8.52874 \times 10^{-2} $ & $  5.23675 \times 10^{-3} $ & \multicolumn{1}{c}{---} \\
& $n = 4$ & $  2.05350 \times 10^{1} $ & $ -8.88641 \times 10^{-2} $ & $  4.72354 \times 10^{-3} $ & $  1.70230 \times 10^{-3} $ \\
\hline
\multicolumn{1}{l}{$\mu^{12}$C} &
$n = 1$ & $  2.82330 \times 10^{1} $ & \multicolumn{1}{c}{---} & \multicolumn{1}{c}{---} & \multicolumn{1}{c}{---} \\
& $n = 2$ & $  2.37752 \times 10^{1} $ & $ -8.39889 \times 10^{-2} $ & \multicolumn{1}{c}{---} & \multicolumn{1}{c}{---} \\
& $n = 3$ & $  2.22857 \times 10^{1} $ & $ -9.66721 \times 10^{-2} $ & $  5.63452 \times 10^{-3} $ & \multicolumn{1}{c}{---} \\
& $n = 4$ & $  2.16554 \times 10^{1} $ & $ -1.00070 \times 10^{-1} $ & $  4.69932 \times 10^{-3} $ & $  2.28402 \times 10^{-3} $ \\
\hline
\multicolumn{1}{l}{$\mu^{13}$C} &
$n = 1$ & $  2.82396 \times 10^{1} $ & \multicolumn{1}{c}{---} & \multicolumn{1}{c}{---} & \multicolumn{1}{c}{---} \\
& $n = 2$ & $  2.37808 \times 10^{1} $ & $ -8.40327 \times 10^{-2} $ & \multicolumn{1}{c}{---} & \multicolumn{1}{c}{---} \\
& $n = 3$ & $  2.22906 \times 10^{1} $ & $ -9.67187 \times 10^{-2} $ & $  5.63562 \times 10^{-3} $ & \multicolumn{1}{c}{---} \\
& $n = 4$ & $  2.16599 \times 10^{1} $ & $ -1.00116 \times 10^{-1} $ & $  4.69868 \times 10^{-3} $ & $  2.28645 \times 10^{-3} $ \\
\hline
\hline
\end{tabular}
\end{table*}
\endgroup

%
%
\subsection{High--Energy Part}
\label{HEP}

In the spirit of NRQED$_\mu$, the 
high-energy effects are described by 
the effective operators given in 
Eq.~\eqref{NRQEDmu}.
From Eqs.~\eqref{NRQEDmu} and~\eqref{matching_coeff},
we obtain two effective potentials $V_1$ and $V_2$.
The first one, $V_1$, is proportional to 
the matching coefficient $c^{(\mu)}_D $.
After the subtraction of the tree-level term,
it reads as follows,
\begin{align}
V_1 &= -[c^{(\mu)}_D - 1] \, \frac{e}{2 m_\mu^2}
\left ( \vec D \cdot \vec E - \vec E \cdot \vec D \right )
\cr
&= -[c^{(\mu)}_D - 1] \, \frac{1}{2 m_\mu^2}
[\vec \nabla, [-\vec \nabla, (V_C + V_\vp)]] 
\cr
&= \frac{\alpha}{3 \pi} \, 
\left \{ -\frac{1}{2 \epsilon}
+ \ln \left ( \frac{m_\mu}{\dimmu} \right ) \right \} 
\frac{1}{m_\mu^2} \vec \nabla^2 ( V_C + V_\vp) \,.
\end{align}
It gives rise to an energy shift
\begin{multline}
E_1 = \langle \Psi_n | V_1 | \Psi_n \rangle 
= \frac{\alpha}{3 \pi} \,
\left \{ -\frac{1}{2 \epsilon}
+ \ln \left ( \frac{m_\mu}{\dimmu} \right ) \right \}
\\
\times \frac{1}{m_\mu^2} 
\langle \Psi_n | \vec \nabla^2 ( V_C + V_\vp) | \Psi_n \rangle\,.
\end{multline}
Upon expansion of the matrix element of $\Psi_n$,
one obtains
\begin{multline}
\label{E1res}
E_1 = 
\frac{4\alpha}{3\pi} \frac{(Z\alpha)^4}{n^3} \,
\frac{m_r^3}{m_\mu^2} \,
\delta_{\ell 0} \left \{ -\frac{1}{2 \epsilon}
+ \ln \left ( \frac{m_\mu}{\dimmu} \right ) \right \} 
\\
+ \left( \frac{\alpha}{\pi} \right)^2 
\frac{(Z\alpha)^4}{n^3} \frac{m_r^3}{m_\mu^2} \,
\left \{ -\frac{1}{2 \epsilon}
+ \ln \left ( \frac{m_\mu}{\dimmu} \right ) \right \} 
V_{61} \,.
\end{multline}
The second effective potential which follows
from Eq.~\eqref{NRQEDmu} and~\eqref{matching_coeff},
is proportional to $c^{(\mu)}_S$ (we exclude the 
tree-level term),
\begin{align}
V_2 &= -[c^{(\mu)}_S - 1] \, 
\frac{\ii \, e}{8 m_\mu^2} \vec \sigma \cdot
\left ( \vec D \times \vec E - \vec E \times \vec D \right )
\cr
&= [c^{(\mu)}_S - 1] \, \frac{\ii \, e}{4 m_\mu^2} \vec \sigma \cdot
\vec E \times \vec \nabla
\cr
&= - [c^{(\mu)}_S - 1] \, \frac{1}{4 m_\mu^2} \vec \sigma \cdot
(-\vec\nabla(V_C + V_\vp)) \times (-\ii \vec \nabla)
\cr
&=  \frac{\alpha}{\pi} \, \frac{1}{4 m_\mu^2} \,
\frac{1}{r} \frac{\partial (V_C + V_\vp)}{\partial r} \,
\vec \sigma \cdot \vec L \,,
\end{align}
where $\vec L =  \vec r \times \vec p $.
Its expectation value, 
$E_2 = \langle \Psi_n | V_2 | \Psi_n \rangle $,
needs to be evaluated. 
For the angular part, we have the 
identity
\begin{equation}
\label{angpart}
\langle \vec \sigma \cdot \vec L \rangle =
-(\kappa + 1) = 
\left\{ \begin{array}{cc}
\ell & \qquad \mbox{$(j = \ell + 1/2)$} \\
-\ell -1 & \qquad \mbox{$(j = \ell - 1/2)$}
\end{array} \right. \,,
\end{equation}
where $\kappa = (-1)^{j+\ell+1/2} \, (j+1/2)$ 
is the Dirac angular quantum number.
Here, we seek the evaluation of the diagonal 
matrix element of a reference 
state with quantum numbers $n$, $\ell$ and $j$
(principal, orbital angular momentum and total angular 
momentum). One also verifies that, for $\ell \neq 0$,
one has the relation
$\langle \vec \sigma \cdot \vec L \rangle/(\ell (\ell+1)) = -1/\kappa$.
For $\ell \neq 0$, one therefore has the relation 
$\left< \phi_n \left| \frac{\vec\sigma \cdot \vec L}{r^3} \right| \phi_n
\right> = \frac{2 (Z\alpha m_r)^3 \langle \vec\sigma \cdot \vec L \rangle}%
{n^3\,\ell\,(\ell+1)\,(2 \ell+1)} = -\frac{2 (Z\alpha m_r)^3}{n^3\,\kappa \,
(2 \ell+1)}$, and we need the matrix element
$\left< \phi_n \left| \frac{1}{r^3} \right| \phi_n  \right> =
\frac{2 (Z\alpha m_r)^3}{n^3\,\ell\,(\ell+1)\,(2 \ell+1)} =
-\frac{2 (Z\alpha m_r)^3}{n^3\,\kappa \,(2 \ell+1)}$.
Finally, $E_2$ evaluates to the following expression,
\begin{align}
\label{E2res}
E_2 = \langle \Psi_n | V_2 | \Psi_n \rangle &=
\frac{\alpha}{\pi} \, \frac{1}{4 m_\mu^2} \,
\left< \phi_n \left| \frac{1}{r} \frac{\partial V_C}{\partial r} \,
\vec \sigma \cdot \vec L \right| \phi_n \right>
\cr
& ~ + \left( \frac{\alpha}{\pi} \right)^2 \, 
\frac{(Z\alpha)^4}{n^3} \, \frac{m_r^3}{m_\mu^2} M_{60} 
\cr
&= \frac{\alpha}{\pi} \, \frac{(Z\alpha)^4}{n^3} \,
\frac{m_r^3}{m_\mu^2} \,
\left(-\frac{1 - \delta_{\ell 0}}{2 \kappa (2 \ell + 1)} \right)
\cr
& ~ + \left( \frac{\alpha}{\pi} \right)^2 \,
\frac{(Z\alpha)^4}{n^3} \, \frac{m_r^3}{m_\mu^2} 
\langle \vec \sigma \cdot \vec L \rangle \, N_{60} \,.
\end{align}
Here,
$M_{60} = \langle \vec \sigma \cdot \vec L \rangle \, N_{60} 
= - ( \kappa + 1) \, N_{60}$
is a coefficient that vanishes for $S$ states.
(The notation $M_{60}$ was used 
in Ref.~\cite{WuJe2011}.) For $N_{60}$, we obtain the result
\begin{multline}
N_{60} = \frac{n^3}{4 (\alpha/\pi) (Z\alpha)^4 m_r^3}
\\
\times
\Biggl(
\left< \phi_n \left| \frac{1}{r} \, \frac{\partial V_{\rm vp}}{\partial r}
\right| \phi_n \right>
+ 2 \, \left< \phi_n \left| \frac{1}{r} \, \frac{\partial V_C}{\partial r} \,
\right| \delta \phi_n \right> \Biggr) \,.
\end{multline}

%
%
\subsection{End Result}

We now add the results from
Eqs.~\eqref{ELEPres},~\eqref{E1res} and~\eqref{E2res},
and observe that both the dependence
on the dimensional parameter $\epsilon$ 
as well as the dependence on the renormalization
scale $\mbox{\dimmu}$ cancel. One obtains two terms,
the first of which is the plain bound-state 
self-energy, while the second one is the 
vacuum-polarization correction to the 
self energy,
\begin{equation}
E_{\rm LEP} + E_1 + E_2 =  
E_{\rm SE} + E_{\rm SE-eVP} \,.
\end{equation}
The plain self-energy is verified as follows,
\begin{multline}
E_{\rm SE} = \frac{\alpha}{\pi} \,
\frac{(Z\alpha)^4}{n^3} \,
\frac{m_r^3}{m_\mu^2} \,
\biggl[ \frac43 \left\{ 
\ln\left( \frac{m_\mu}{ (Z\alpha)^2 m_r } \right) 
+ \frac56 \right\} \,
\delta_{\ell 0} 
\\
- \frac{m_\mu}{m_r} \frac{1 - \delta_{\ell 0}}{2 \kappa (2 \ell + 1)} 
- \frac43 \ln k_0(n, \ell) \biggr] \,.
\end{multline}
We here restore the correct
reduced-mass dependence of the 
anomalous-magnetic-moment term, 
which is due to the proton's convection current~\cite{JeAd2022book}.
The end result for the SE-eVP term is as follows,
\begin{multline}
\label{master}
E_\mathrm{SE-eVP} =
\left( \frac{\alpha}{\pi} \right)^2 \, 
\frac{(Z\alpha)^4}{n^3} \, \frac{m_r^3}{m_\mu^2} \,
\biggl[ V_{61} \,
\biggl\{ \ln\left( \frac{m_\mu}{(Z\alpha)^2 m_r} \right)
\\
+ \frac{5}{6} \biggr\}
+ \frac{m_\mu}{m_r} \, 
\langle \vec \sigma \cdot \vec L \rangle \, N_{60} 
- \frac43 \, L_{60} \biggr] \,.
\end{multline}
The angular part $\langle \vec \sigma \cdot \vec L \rangle$ 
is given in Eq.~\eqref{angpart}.

For reference, we may point out that, in comparison to the treatment in
Ref.~\cite{Je2011aop1}, the final result is to eliminate the ``2'' in the
argument of the logarithm, replace $10/9 \to 5/6$, and supply prefactor
$m_\mu/m_r$ in front of $M_{60}$.  We also take the opportunity to correct the
reduced mass dependence of the spin-orbit effect as compared to Ref.~\cite{WuJe2011},
namely, the presence of the additional factor $m_\mu/m_r$.

\def\arraystretch{1.2}

\begin{table}[t!] 
\caption{
\label{tab:lead}
Comparison of leading-logarithmic results for the SE-eVP 
correction to various quantities, in meV.
The notation for the energy intervals is $nL$--$nL'\equiv E(nL)-E(nL')$.
By the leading logarithmic approximation, we understand the result obtained
by setting $N_{60} = L_{60} = 0$ in Eq.~\eqref{master}.
We note that the inclusion of the nonlogarithmic 
terms shifts the result considerably. 
For example, in the case of $\mu {}^{11}{\rm B}$,
the result for $2S$--$1S$ shifts from $-57.4$\,meV
to $-26.4$\,meV.}
\begin{tabular}{l,.l}
\hline
\hline
& \multicolumn{1}{c}{This Work} & \multicolumn{2}{c}{Others} \\
\hline
$2S$    in $\mu$H      & 0.x00488  &   0.x00486 & (Ref.~\cite{Bo2012}) \\
$2S$    in $\mu$D      & 0.x00586  &   0.x00589 & (Ref.~\cite{Bo2012}) \\
$2S$    in $\mu$$^3$He & 0.x1273   &   0.x1277  & (Ref.~\cite{Bo2012}) \\
$2S$    in $\mu$$^4$He & 0.x1313   &   0.x1314  & (Ref.~\cite{Bo2012}) \\
\hline
$2P$--$2S$ in $\mu$H      & -0.x00492 &   -0.x005   & (Ref.~\cite{Pa1996muonic}) \\
                     &           &   -0.x00490 & (Ref.~\cite{Bo2012}) \\
$2P$--$2S$ in $\mu$D        & -0.x00591  & -0.x0047 & (Ref.~\cite{KrMa2011}) \\
$2P$--$2S$ in $\mu$$^3$He   & -0.x1285   & -0.x1008 & (Ref.~\cite{KrMaMaFa2015}) \\
$2P$--$2S$ in $\mu$$^4$He   & -0.x1325   & -0.x1074 & (Ref.~\cite{KrMaMaFa2015}) \\
$2P$--$2S$ in $\mu$$^6$Li   & -0.x764    & -0.x23   & (Ref.~\cite{KrMaMaSu2016}) \\
$2P$--$2S$ in $\mu$$^7$Li   & -0.x771    & -0.x23   & (Ref.~\cite{KrMaMaSu2016}) \\
$2P$--$2S$ in $\mu$$^9$Be   & -2.x60     & -0.x71   & (Ref.~\cite{KrMaMaSu2016}) \\
$2P$--$2S$ in $\mu$$^{10}$Be& -2.x61     & -0.x71   & (Ref.~\cite{KrMaMaSu2016}) \\
$2P$--$2S$ in $\mu$$^{10}$B & -6.x61     & -1.x66   & (Ref.~\cite{KrMaMaSu2016}) \\
$2P$--$2S$ in $\mu$$^{11}$B & -6.x63     & -1.x66   & (Ref.~\cite{KrMaMaSu2016}) \\
\hline
$2S$--$1S$ in $\mu$H      & -0.x0392 & -0.x0281 & (Ref.~\cite{DoEtAl2019}) \\
$2S$--$1S$ in $\mu$D      & -0.x0473 & -0.x0285 & (Ref.~\cite{DoEtAl2019}) \\
$2S$--$1S$ in $\mu$$^3$He & -1.x089  & -0.x2597 & (Ref.~\cite{DoEtAl2019}) \\
$2S$--$1S$ in $\mu$$^4$He & -1.x123  & -0.x2708 & (Ref.~\cite{DoEtAl2019}) \\
$2S$--$1S$ in $\mu$$^7$Li   & -6.x66 & -1.x93   & (Ref.~\cite{DoEtAl2021}) \\
$2S$--$1S$ in $\mu$$^9$Be   & -22.x6 & -6.x21   & (Ref.~\cite{DoEtAl2021}) \\
$2S$--$1S$ in $\mu$$^{11}$B & -57.x4 & -15.x18  & (Ref.~\cite{DoEtAl2021}) \\
\hline
$3S$--$1S$  in $\mu$H        & -0.x0427 & -0.x0294 & (Ref.~\cite{DoEtAl2020}) \\
\hline
\hline
\end{tabular}
\end{table}

%
%
\section{Numerical Results}
\label{sec3}

\subsection{Numerical Methods}
\label{sec3A}

Let us discuss the numerical methods
employed in the evaluation of both 
the low-energy part (Sec.~\ref{LEP})
and the high-energy part (Sec.~\ref{HEP}).

The computationally easiest task is to evaluate the 
correction $L_{60}^{(E)}$ given in Eq.~\eqref{eq:LE}; the procedure
is completely analogous to the relativistic-energy
correction to the Bethe logarithm 
given in Eq.~(42) of Ref.~\cite{JePa1996}.
One calculates the derivative (with respect to the 
reference-state energy) of the 
matrix element of the (nonrelativistic) dynamic polarizability 
of the reference state. 
For $1S$, $2S$ and $3S$ states, analytic results
for the dynamic polarizability 
have been given in Refs.~\cite{Pa1993,AdCaArJe2022}.
After differentiation with respect to the 
reference-state energy, one multiplies by the 
diagonal matrix element of the Uehling potential,
integrates over the energy of the 
virtual photon up to an upper cutoff $\Lambda$,
and extracts the finite part of the 
integral according to the procedure
outlined in Eqs.~\eqref{ELEPstart}---\eqref{CANCEL}.

An alternative method for the 
calculation of correction to the 
Bethe logarithm due to the reference-state energy 
is based on a discretization of the 
Schr\"{o}dinger--Coulomb problem
on an exponential lattice~\cite{SaOe1989,PaSiZa2017}
where one obtains a pseudo-spectrum 
representing the continuum states.
This method has recently been used~\cite{WuJe2008}
for the calculation of relativistic 
Bethe logarithms for highly excited $D$ states
in hydrogenlike systems.
The advantage of the latter method is 
that unified formulas can be used 
for the evaluation of the transition matrix 
elements of the reference to the virtual state,
and for the summation over the 
virtual continuum states.

The second method, based on a discretized
representation of the Schr\"{o}dinger--Coulomb propagator,
has computational advantages for the 
wave-function correction $L_{60}^{(\psi)}$,
as given in Eq.~\eqref{eq:LP}, and
especially for the Hamiltonian correction $L_{60}^{(H)}$,
given in Eq.~\eqref{eq:LH}.
In the analytic approach,
if one inserts the Uehling potential as a 
perturbation to the dynamic polarizability
as in Eq.~(41) of Ref.~\cite{JePa1996},
one invariably ends of with matrix elements involving
two Schr\"{o}dinger--Coulomb
propagators with the Uehling potential sandwiched 
in between. The calculation necessitates a 
double summation over the virtual states 
of the Sturmian decomposition of the 
Schr\"{o}dinger--Coulomb propagators.
It is computationally a lot easier to evaluate,
explicitly and on an exponential numerical 
lattice~\cite{SaOe1989,WuJe2008}, the 
sums over the virtual states given in Eq.~\eqref{eq:LH}.

Numerical results for $V_{61}$, $N_{60}$ and $L_{60}$ are given in Tables~\ref{tableV61},~\ref{tableN60} and~\ref{tableL60}, respectively.  These
can be used in order to evaluate the SE-eVP correction for all reference states
with principal quantum numbers $n \leq 4$,  for muonic ions with principal
quantum numbers $Z \leq 6$, as indicated.  Our numerical results for $V_{61}$
are more accurate than, and confirm the entries in, Eq.~(3.8) of
Ref.~\cite{Je2011aop1} where a nonperturbative approach in the
vacuum-polarization potential was employed.

%
%
\subsection{Leading Logarithm}
\label{sec3B}

The energy proportional to $V_{61}$ in Eq.~\eqref{master} corresponds to the
leading-logarithmic approximation which was calculated a while ago for the Lamb
shift in muonic hydrogen by Pachucki~\cite{Pa1996muonic}. As several authors
have used this approximation, it is instructive to compare our results with
theirs in Table~\ref{tab:lead}.  We obtain a good agreement with the original
calculation by Pachucki (Ref.~\cite{Pa1996muonic}), and Borie's
calculations~\cite{Bo2012}, and some of the results communicated in
Ref.~\cite{KrMaMaFa2015}, but discrepancies remain with respect to other
investigations.  The origin of the disagreement is not clear to us.  

%
%
\subsection{Lamb Shift}
\label{sec3C}

While the numerical results for $V_{61}$, $N_{60}$ and $L_{60}$,
given in Tables~\ref{tableV61},~\ref{tableN60} and~\ref{tableL60}, 
allow us to evaluate the
SE-eVP correction for all reference states
with principal quantum numbers $n \leq 4$,
it is instructive to present a few
numerical examples. We choose the 
$2P_{1/2}$--$2S$ energy difference, and remember that,
according to Eq.~\eqref{master}, there is a 
residual dependence on the electron-spin orientation.
We obtain the following results,
\begin{subequations}
\begin{align}
\calL(\mu\mbox{H}) =& \;           -0.002715 \, {\rm meV} \,, \\
\calL(\mu\mbox{D}) =& \;           -0.003267 \, {\rm meV} \,, \\
\calL(\mu {}^3 \mbox{He}) =& \;    -0.067312 \, {\rm meV} \,, \\
\calL(\mu {}^4 \mbox{He}) =& \;    -0.069711 \, {\rm meV} \,, \\
\calL(\mu {}^6 \mbox{Li}) =& \;    -0.381975 \, {\rm meV} \,, \\
\calL(\mu {}^7 \mbox{Li}) =& \;    -0.385313 \, {\rm meV} \,, \\
\calL(\mu {}^9 \mbox{Be}) =& \;    -1.23960 \, {\rm meV} \,, \\
\calL(\mu {}^{10} \mbox{Be}) =& \; -1.24460 \, {\rm meV} \,, \\
\calL(\mu {}^{10} \mbox{B}) =& \;  -2.99936\, {\rm meV} \,, \\
\calL(\mu {}^{11} \mbox{B}) =& \;  -3.00906 \, {\rm meV} \,, \\
\calL(\mu {}^{12} \mbox{C}) =& \;  -6.10714 \, {\rm meV} \,, \\
\calL(\mu {}^{13} \mbox{C}) =& \;  -6.12101 \, {\rm meV} \,.
\end{align}
\end{subequations}
Numerically, these entries are a lot smaller in 
magnitude than those obtained in the leading-logarithmic
approximation from Table~\ref{tab:lead}, in view of a partial mutual 
cancelation between the logarithmic and non-logarithmic terms.

Our result for the $2P_{1/2}$--$2S$ Lamb shift in muonic hydrogen differs by less than $1\%$ from that of a recent calculation, which employed different numerical methods~\cite{PaSiZa2017}. Nevertheless, the contributions to individual levels differ by up to $25\%$.

%
%
\section{Conclusions}
\label{sec4}

In this paper, we have evaluated the 
combined self-energy vacuum-polarization
correction to the bound-state energy 
levels of one-muon ions with nuclear
charge numbers $Z \leq 6$. 
We have investigated the one-muon ions with stable (or very long-lived) nuclei,
i.e., $\mu\mbox{H}$ and $\mu\mbox{D}$ ($Z=1$), 
$\mu {}^3 \mbox{He}$ and
$\mu {}^4 \mbox{He}$ ($Z=2$), 
$\mu {}^6 \mbox{Li}$ and
$\mu {}^7 \mbox{Li}$ (with $Z=3$),
$\mu {}^9 \mbox{Be}$ and
$\mu {}^{10} \mbox{Be}$ (with $Z=4$),
$\mu {}^{10} \mbox{B}$ and
$\mu {}^{11} \mbox{B}$ (where $Z=5$),
and carbon ions, $\mu {}^{12} \mbox{C}$ and 
$\mu {}^{13} \mbox{C}$ ($Z=6$).

According to Eq.~\eqref{master},
the SE-eVP correction can be expressed in 
terms of three coefficients, $V_{61}$,
$N_{60}$ and $L_{60}$,
which represent the leading-logarithmic 
approximation ($V_{61}$),
the anomalous-magnetic-moment term ($N_{60}$),
and the eVP-induced correction to the 
Bethe logarithm ($L_{60}$).
Results are given in Tables~\ref{tableV61},~\ref{tableN60}
and~\ref{tableL60}. In light muonic atoms, we confirm that
the SE-eVP effect is the largest out
of all corrections which enter at 
$\alpha^2 (Z\alpha)^4 m_r$~\cite{KaIvKa2013},
as illustrated in Fig.~\ref{fig:cont}.

The numerical methods used in our 
calculations are described in Sec.~\ref{sec3A}.
We rely on a discretization of the 
Schr\"{o}dinger--Coulomb propagator 
on a numerical lattice~\cite{SaOe1989,WuJe2008,PaSiZa2017}.
The discrete-state representations
of the energy, wave-function and 
Hamiltonian corrections to the Bethe logarithm
are employed [see Eqs.~\eqref{eq:LE},~\eqref{eq:LP}
and~\eqref{eq:LH}].
Values for the contributions to the Lamb
shift (in the leading logarithmic approximation)
are given in Sec.~\ref{sec3B},
while the nonlogarithmic terms are added
in Sec.~\ref{sec3C}.
A partial mutual cancelation is observed between the logarithmic and nonlogarithmic 
terms. Our calculations advance the precision theory 
of the spectrum of muonic ions.

\section*{Acknowledgments}

The authors acknowledge helpful conversations
with Prof.~Gregory~S.~Adkins.
Support from the National Science Foundation
(grant PHY--2110294) is gratefully acknowledged.
B.~O.~is grateful for the support of the Council for 
Higher Education Program for Hiring Outstanding Faculty 
Members in Quantum Science and Technology.

\appendix

%
%
\section{Uehling Potential and Numerical Integration}
\label{appa}

We aim to find a representation
of the Uehling potential suitable
for numerical integration on a lattice.
To this end, we start from Eq.~(10.245) of Ref.~\cite{JeAd2022book},
where the Uehling potential is given as follows,
\begin{equation}
\label{qrep}
V_\vp(r) = -\frac{2 \alpha (Z\alpha)}{3 \pi r}
\int_{2m}^\infty \dd q \, \frac{\ee^{-q r}}{q} 
\sqrt{ 1 - \frac{4 m^2}{q^2} }
\left( 1 +  \frac{2 m^2}{q^2} \right).
\end{equation}
We are integrating from the threshold $q = 2m$
of pair production.
Now, we should transform to atomic units,
adapted to a two-particle muonic bound system
with reduced mass $m_r$,
\begin{align}
\beta =& \; \frac{m_e}{Z \alpha m_r} = a_0 \, m_e \,,
& a_0 & = \frac{1}{Z \alpha m_r} \,,
\\
r =& \; a_0 \rho \,,
& m_e r & = \beta \rho \,.
\end{align}
This defines the generalized Bohr radius $a_0$ and 
the electron-nucleus distance
$\rho$ in atomic units. 
Recoil parameters for muonic bound systems 
with nuclear charge numbers $1 \leq Z \leq 6$
are as follows,
$\beta(\mu\mbox{H}) = 0.737384$,
$\beta(\mu\mbox{D}) = 0.700086$,
$\beta(\mu {}^3 \mbox{He}) = 0.343843$,
$\beta(\mu {}^4 \mbox{He}) = 0.340769$,
$\beta(\mu {}^6 \mbox{Li}) = 0.225085$,
$\beta(\mu {}^7 \mbox{Li}) = 0.224491$,
$\beta(\mu {}^9 \mbox{Be}) = 0.167774$,
$\beta(\mu {}^{10} \mbox{Be}) = 0.167565$, 
$\beta(\mu {}^{10} \mbox{B}) = 0.134052$,
$\beta(\mu {}^{11} \mbox{B}) = 0.133916$,
$\beta(\mu {}^{12} \mbox{C}) = 0.111503$,
$\beta(\mu {}^{13} \mbox{C}) = 0.111422$.
After the substitution
$q = 2 m \left[  1 + \frac{\xi}{2 \beta \rho}\right]$,
the scaled Uehling potential $U_\vp(\rho)$,
for a muonic bound system with recoil parameter $\beta$,
is obtained as 
\begin{multline}
\label{Uvp}
U_\vp(\rho) = \frac{V_\vp(r)}{(\alpha/\pi) \, (Z\alpha)^2 m_r} 
= -\frac{2 \exp(-2 \beta \rho)}{3 \rho} 
\\
\times \int_0^\infty  \dd \xi \, \ee^{-\xi} \,
\sqrt{\xi} \, \sqrt{4 \beta \rho + \xi} \,
\frac{6 (\beta \rho)^2 + 4 \beta \rho \xi + \xi^2}{(2 \beta \rho + \xi)^4} \,.
\end{multline}
In the numerical implementation, one integrates over $\xi$
from $\xi = 0$ to $\xi = 4$ by Gauss--Legendre integration,
then using Gauss--Laguerre from $\xi = 4$ to $\xi = \infty$.

\end{document}